\documentclass[9pt,conference]{IEEEtran}

\usepackage[preprint]{waspaa25}

\usepackage{bm} 

\usepackage{xcolor}
\usepackage{cite}
\usepackage{siunitx}
\usepackage{amssymb, amsmath, amsfonts}
\usepackage{subfigure}
\usepackage{url}

\usepackage{tikz}
\usetikzlibrary{positioning}
\usepackage{graphicx}
\makeatother
\usepackage{pgfplots}%
	\usepgfplotslibrary{units}
	\usepgfplotslibrary{groupplots}%
	\pgfplotsset{compat=1.9}
\makeatletter
\pgfplotsset{
unit code/.code 2 args=
    \begingroup
    \protected@edef\x{\endgroup\si{#2}}\x
}

\newcommand{\NMSE}{\operatorname{NMSE}}

\title{Low-Rank Adaptation of Deep Prior Neural Networks For Room Impulse Response Reconstruction}

\name{Mirco Pezzoli$^{1}$,
      Federico Miotello$^{1}$,
      Shoichi Koyama$^{2}$,
      Fabio Antonacci$^{1}$\thanks{}}
\address{$^{1}$Politecnico di Milano, Milan, Italy \;
$^{2}$National Institute of Informatics, Tokyo, Japan\\
}

\begin{document}

\maketitle

\begin{abstract}
The Deep Prior framework has emerged as a powerful generative tool which can be used for reconstructing sound fields in an environment from few sparse pressure measurements.
It employs a neural network that is trained solely on a limited set of available data and acts as an implicit prior which guides the solution of the underlying optimization problem.
However, a significant limitation of the Deep Prior approach is its inability to generalize to new acoustic configurations, such as changes in the position of a sound source.
As a consequence, the network must be retrained from scratch for every new setup, which is both computationally intensive and time-consuming.
To address this, we investigate transfer learning in Deep Prior via Low‑Rank Adaptation (LoRA), which enables efficient fine-tuning of a pre-trained neural network by introducing a low-rank decomposition of trainable parameters, thus allowing the network to adapt to new measurement sets with minimal computational overhead.
We embed LoRA into a MultiResUNet‑based Deep Prior model and compare its adaptation performance against full fine-tuning of all parameters as well as classical retraining, particularly in scenarios where only a limited number of microphones are used.
The results indicate that fine-tuning, whether done completely or via LoRA, is especially advantageous when the source location is the sole changing parameter, preserving high physical fidelity, and highlighting the value of transfer learning for acoustics applications.

\end{abstract}

\section{Introduction}
Sound field reconstruction (SFR) aims at estimating the acoustic pressure field over an extended spatial region from a limited set of microphone measurements. This task is crucial in numerous applications that range from augmented and virtual reality, where immersive audio is based on accurate reproduction of room acoustics~\cite{vorlander2015virtual}, to teleconferencing and active noise control. 
At the core of SFR lies the estimation of room impulse responses (RIRs), which fully characterize the propagation between a source and each sensor in a given environment. 
RIR estimation enables downstream tasks such as source localization~\cite{cobos2017survey,cobos2023acoustic}, separation~\cite{gannot2017consolidated},  imaging \cite{olivieri2024acoustic} and immersive rendering \cite{mccormack2022parametric}.


Driven by advances in deep learning in acoustics \cite{van2025deep}, deep learning methods have emerged as powerful alternatives to SFR.
Supervised Convolutional Neural Networks (CNN) models, such as the U‑Net inpainting strategy proposed in \cite{lluis2020sound} demonstrated that missing room transfer functions can be recovered in the frequency domain, but their reliance on large, diverse datasets limits generalization to unseen rooms and setup conditions. 
Alternatively, physics-informed neural networks (PINNs)~\cite{raissi2019physics,cuomo2022scientific,koyama2025physics} incorporate partial differential equation constraints, i.e., Helmholtz or wave equations, into the loss function~\cite{shigemi2022physics, olivieri2024physics,bi2024point, karakonstantis2024room,ma2023physics}, reducing the need for training data while improving physical consistency. 
In fact, the generalization limits of PINN are bypassed by its per-element training strategy that, however, requires a full retraining for each new acoustic configuration.
In parallel, generative frameworks have been investigated in the literature in order to address SFR. 
In \cite{fernandez2023generative,karakonstantis2023generative} GANs have been adopted for the reconstruction of RIRs also in combination with physical priors. 
More recently, denoising diffusion probabilistic models (DDPMs) \cite{ho2020denoising} which represent the state-of-the-art generative models in several audio-related tasks \cite{lemercier2025diffusion, moliner2024buddy,sanchez2025towards,thuillier2025hrtf,juanpere2024diffusion} has been applied to RIR estimation. 
In \cite{miotello2024reconstruction}, DDPM is proposed in order to reconstruct the magnitude of low-frequency acoustic fields. 
Later, in \cite{lluis2025blind, della2025diffusionrir} DPPMs are updated to solve SFR directly in time domain. 
While GANs and DDPMs can synthesize realistic RIRs, they depend on large datasets for offline training and thus their capability to generalize to unseen scenarios is not fully explored yet. 
In contrast, the Deep Prior (DP) framework~\cite{ulyanov2018deep} represents itself a generative method that, similarly to PINNs, fits a network per element without additional training data sets, namely, using only the observed sparse measurements. 
The DP method has been employed in different applications \cite{kong2020deep, malvermi2021frfdip, miotello2023deep,hendriksen2020noise2inverse}, including the reconstruction of RIRs in a uniform linear arrays (ULAs) setup~\cite{pezzoli2022deep}.
This approach can lead to SFR with minimal data at the cost of retraining the entire network from scratch for every new room or sensor layout.

Despite their strengths, all the above data‐driven methods share a common limitation: their generalization to unknown acoustic scenarios is unclear.
On the one hand, retraining a model from scratch for each new acoustic scenario or configuration might be unpractical. 
On the other hand, we often have access to previous recordings of one or more environments with similar acoustic properties of the one we aim at reconstructing, and intuitively we can expect that a model pretrained on one room or microphone configuration can be effectively adapted to another.
This motivates the need for efficient fine-tuning techniques to adapt pretrained models to a new setup with minimal computational overhead and limited data.

The interest in adapting, or fine-tuning, existing architectures has increased in the recent years especially driven by the adoption of large language models and other large‑scale generative neural networks \cite{lu2025fine} leading also to the introduction of a challenge on data generation\footnote{\url{https://sites.google.com/view/genda2025}}. 
In this context, \textit{Low-Rank Adaptation} (LoRA)~\cite{hu2022lora} appeared as an efficient and effective approach for adapting pre-trained models to different tasks and perform transfer learning. 
This adaptation method allows one to efficiently fine-tune pretrained neural networks by injecting trainable low-rank matrices into existing weights, without altering the original network parameters.
This possibility is interesting in order to avoid retraining the models from scratch for every new environment or configuration, in particular, because DPs typically have larger networks with respect to PINNs. 
To the best of our knowledge, there exists limited study of fine‐tuning strategies for DP or other deep-learning SFR frameworks. 
Although LoRA has been successfully applied in the context of HRTF personalization \cite{masuyama2024niirf} and synthesis of Ambisonics signals \cite{ick2025direction}.

In this work, we therefore aim at analyzing the efficacy of adaptation in the DP context.
We investigate the integration of LoRA into a MultiResUNet‐based DP model \cite{pezzoli2022deep} in order to efficiently adapt the model to a new set of measurements by reusing a pretrained architecture and updating only a small number of parameters.
We compare the performance of LoRA adaptation with respect to a full fine-tuning of the architecture (FT) and the classical DP retraining when a limited number of microphones are available. 
The results highlight that fine-tuning, either full or LoRA, is beneficial especially when the source position is the only parameter that changes between configurations. 
When adaptation to different rooms is considered, the performance is in line with a training from scratch, although LoRA uses around the $\SI{30}{\percent}$ of the trainable parameters.
This suggests that LoRA can be particularly useful in scenarios where only a few microphone measurements are available or where computational resources are limited, since only a small number of parameters has to be trained and swapped within the architecture.
\vspace{0em}
\section{Proposed Method}\label{sec:method}
\subsection{Signal Model}
Let us consider an acoustic environment where a source emits a signal $s(t)$ at position $\bm{r}' = [x',y',z']^T$, and a set of microphones is placed at positions $\{\bm{r}_m\}_{m=1}^M$. Under the assumption of Linear Time Invariant (LTI) system, the recorded signal at each microphone can be expressed as the convolution between the source signal and the corresponding RIR
\begin{equation}\label{eq:rir_equation}
    p_m(t) = h_m(\bm{r}', \bm{r}_m, t) * s(t) + e_m(t),
\end{equation}
where $*$ is the linear convolution, $h_m(\bm{r}', \bm{r}_m, t)$ is the RIR between the source and the $m$th microphone, and $e_m(t)$ models additive noise at the sensor.
Given the fact that the RIR $h_m(\cdot)$ completely characterizes the sound field generated by the source in the environment, SFR is cast as the reconstruction of RIRs across the dense spatial grid of $M$ points covering a region of interest.
\subsection{Deep-prior based RIR reconstruction}
Given the sparsely measured RIRs $\tilde{\mathbf{H}}\in\mathbb{R}^{N\times \tilde{M}}$, which is a subset of the $M$ RIRs of $N$ temporal samples to estimate. 
In this context, SFR is described in terms of an optimization problem as 
\begin{equation}\label{eq:rir_problem}
    \hat{\mathbf{H}}^*= \underset{{\hat{\mathbf{H}}}}{\text{argmin}}\, J\left(\hat{\mathbf{H}}, \tilde{\mathbf{H}}\right) =\underset{{\hat{\mathbf{H}}}}{\text{argmin}}\,\left[E\left(\hat{\mathbf{H}}, \tilde{\mathbf{H}}\right) + R\left(\hat{\mathbf{H}}\right)\right],
\end{equation}
where the time index has been omitted for simplicity, $\hat{\mathbf{H}}\in \mathbb{R}^{N\times M}$ are the reconstructed RIRs, $E(\cdot)$ is a data-fidelity term e.g., the mean square error, and $R(\cdot)$ represents a regularizer, required in order to obtain feasible reconstruction.
Different regularization strategies have been proposed in the literature, including sparsity \cite{koyama2019sparse} and physical constraints \cite{ribeiro2023kernel, damiano2025zero}.

Through the DP approach \cite{ulyanov2018deep}, the RIRs are modeled as the output of a CNN with autoencoder structure that takes as input a fixed noise tensor and learns to map it onto a RIR representation.
This generative process aims at avoiding any explicit regularization, assuming that the networks structure itself can serve as a guidance for the correct solution of \eqref{eq:rir_problem}.

Let $\mathcal{N}_{\bm{\theta}}(\mathbf{z})$ be the DP architecture with trainable parameters $\bm{\theta}$ and a random fixed input $\mathbf{z}$. 
The model is trained by minimizing the mismatch between the estimated RIRs at known microphone positions and the measured observation truth
\begin{equation}\label{eq:solution_dp}
    \bm{\theta}^* = \underset{\bm{\theta}}{\text{argmin}}\, E\left( \mathcal{N}_{\bm{\theta}}(\mathbf{z}) \mathbf{S},  \tilde{\mathbf{H}} \right), 
\end{equation}
where $E(\cdot)$ is the $\ell_1$ norm and $\mathbf{S}$ is a sampling operator that selects only the $\tilde{M}$ estimates of the observations \cite{pezzoli2022deep}.
This formulation leverages the implicit bias of CNN structure to generate smooth and coherent spatial structures and thus the explicit regularization in \eqref{eq:rir_problem} is discarded \cite{pezzoli2022deep}.
While the fit is performed over the observation $\tilde{\mathbf{H}}$, the network outputs the RIRs also in \textit{unknown} positions since it reconstructs RIRs for every $M$ location, thus the DP estimate of $h_m$, $m = 1,\dots, M$ is defined as
\begin{equation}
    \hat{\mathbf{H}} = \mathcal{N}_{\bm{\theta}^*}(\mathbf{z}).
\end{equation}
It is worth noting that DP adopts a per-element training and it does not require a training stage as in classical supervised methodologies.
This advantage comes at the cost of repeating the optimization \eqref{eq:solution_dp} when the observation $\tilde{\mathbf{H}}$ changes e.g., due to different source position or new environments.

\subsection{Low-Rank Adaptation of DP Network}
Given a $k\times k$ convolutional layer with $C_{in}$ inputs and $C_{\mathrm{out}}$ channels in $\mathcal{N}_{\bm{\theta}}$, we adapt the layer by replacing the standard convolutional blocks with a LoRA-based formulation \cite{ding2024lora}. 
Specifically, the weights $\mathbf{W} \in \mathbb{R}^{C_{\text{out}} \times C_{\text{in}} \times k \times k}$ are augmented by means of tensor 
\begin{equation}\label{eq:lora}
    \Delta \mathbf{W} = \alpha \cdot \mathbf{B} \mathbf{A},
\end{equation}
where $\mathbf{A}\in\mathbb{R}^{r\times C_{\text{in}}\times k}$ and $\mathbf{B}\in\mathbb{R}^{C_{\text{out}}\times k\times r}$ factorize the weights, and $\alpha$ is a scaling factor that controls the update strength. The tensor multiplication in \eqref{eq:lora} is implemented via a contraction over the low-rank dimension $r$ and the resulting tensor is reshaped into the size of $\mathbf{W}$ \cite{ding2024lora}. 
Thus, the output of adapted convolution is given by 
\begin{equation}\label{eq:adaptation}
    \mathbf{Y}(\mathbf{x}) = (\mathbf{W} + \Delta\mathbf{W}) * \mathbf{x},
\end{equation}
where $\mathbf{x}$ is the input of the layer, and $\mathbf{W}$ are the pretrained weights. 
During the fine-tuning, only $\mathbf{A}$ and $\mathbf{B}$ are updated, while $\mathbf{W}$ is kept frozen. 
This approach enables efficient adaptation with a significant reduction in the number of training parameters \cite{ding2024lora}. 
This LoRA-based strategy allows one to adapt a DP model trained in a given setup to new acoustic environments or source/microphone configurations using very limited data and training parameters. 
Moreover, multiple adaptations can be performed simply by swapping the adapters $\Delta\mathbf{W}$ in \eqref{eq:adaptation}.
\section{Validation}\label{sec:validation}
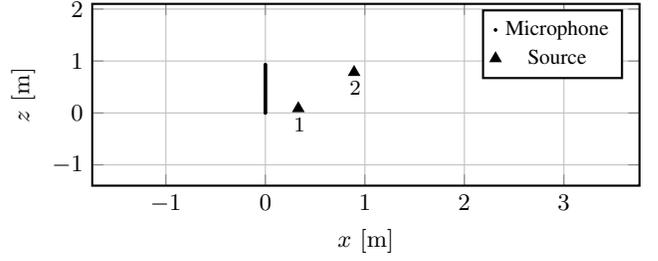
\begin{figure}
    \centering
    \begin{tikzpicture}%
\colorlet{polarcolor}{black}%
\colorlet{srccolor}{black}%
\tikzstyle{source}=[mark = triangle*, draw=srccolor, fill=polarcolor, inner sep=0pt]%
\tikzstyle{mic}=[circle,  mark size = 0.3pt, draw=black, fill=black, inner sep=0pt]%
\begin{axis}[xlabel={$x$}, width=\columnwidth, height=4cm,%
x unit = \meter,%
ylabel={$z$},%
y unit = \meter,%
thick, ymin = -1.4,
ymax=2.1,
xmax=3.76,
xmin= -1.74,
every node near coord/.append style={xshift=3pt, yshift=-6pt, anchor=east,font=\footnotesize},%
enlarge x limits=false,%
enlarge y limits=false,%
grid,legend pos = north east, legend style={transform shape,xshift=0pt, font=\footnotesize}]%
\addplot[only marks, style=mic] file  {images/data/rev_micPos.txt};%
\addlegendentry{Microphone};%
\addplot[only marks,style = source,nodes near coords={\pgfmathprintnumber{\labelz}},
             visualization depends on={value \thisrowno{2}\as\labelz},
             ] table {images/data/reduced_pos.txt};%
\addlegendentry{Source};%
\end{axis}\end{tikzpicture}\vspace{-3em}
    \caption{Geometrical setup of the fixed environment scenario. A ULA of $M=32$ sensors is acquiring the RIRs of two sources with distances of $\approx\SI{30}{\centi\meter}$ and $\SI{90}{\centi\meter}$ (top view).}\label{fig:setup_office}\vspace{-1em}%
\end{figure}
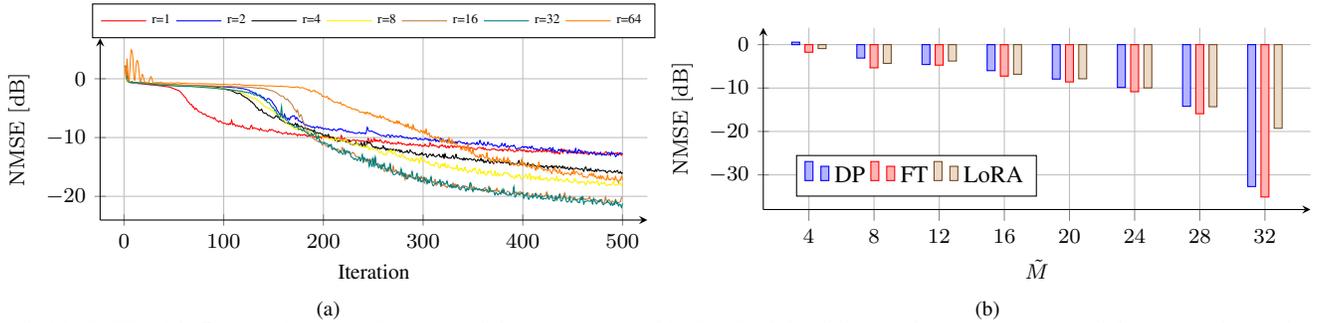
\begin{figure*}[tb]
    \centering
     \begin{subfigure}[]{%
    \begin{tikzpicture}%
\begin{axis}[%
xlabel={Iteration},%
ylabel={$\NMSE$},%
y unit = \decibel, %
axis x line=bottom,%
axis y line=left, %
grid, %
height=4cm, %
width=\columnwidth, %
enlarge x limits=0.05,%
enlarge y limits=0.08,%
style={font=\footnotesize},%
legend columns=7,%
legend style={at={(0.5,1)},anchor=south, font=\tiny},%
cycle list name=color list]%
\addplot+[mark=none, smooth] table {images/data/rank_1.txt};%
\addlegendentry{r=1}%
\addplot+[mark=none, smooth] table {images/data/rank_2.txt};%
\addlegendentry{r=2}%
\addplot+[mark=none, smooth] table {images/data/rank_4.txt};%
\addlegendentry{r=4}%
\addplot+[mark=none, smooth] table  {images/data/rank_8.txt};%
\addlegendentry{r=8}%
\addplot+[mark=none, smooth] table  {images/data/rank_16.txt};%
\addlegendentry{r=16}%
\addplot+[mark=none, smooth] table  {images/data/rank_32.txt};%
\addlegendentry{r=32}%
\addplot+[mark=none, smooth] table  {images/data/rank_64.txt};%
\addlegendentry{r=64}%
\end{axis}%
\end{tikzpicture}\label{fig:rank}}%
    \end{subfigure}%
    \begin{subfigure}[]{%
    \begin{tikzpicture}%
\begin{axis}[
    xlabel={$\tilde{M}$},
    ylabel={$\NMSE$},
    y unit = \decibel,
    axis x line=bottom,
    axis y line=left,
    grid,
    height=4cm,
    width=\columnwidth,
    enlarge x limits=0.1,
    enlarge y limits=0.08,
    style={font=\footnotesize},
    legend columns=4,
    legend style={at={(0.5,0.3)},anchor=north east,font=\normalsize},
    cycle list name=color list,
    ybar,
    bar width=3pt, 
    xtick = data,
    xmax = 32,
    xmin = 4,
    ymax = 1
]
\addplot+[ybar] table {images/data/mic_dip.txt};
\addlegendentry{DP}
\addplot+[ybar] table {images/data/mic_fine.txt};
\addlegendentry{FT}
\addplot+[ybar] table {images/data/mic_lora.txt};
\addlegendentry{LoRA}
\addlegendentry{r=256}
\end{axis}
\end{tikzpicture}\label{fig:mic}}%
    \end{subfigure}%
    \vspace{-1em}\caption{(a) $\operatorname{NMSE}$ of LoRA as a function of iteration of the optimization. (b) $\operatorname{NMSE}$ of the different adaptation strategies while varying the number of observations $\Tilde{M}$.}%
\end{figure*}
\begin{table*}[tb!]
    \resizebox{0.33\textwidth}{0.05\textwidth}{%
    \centering{
    \newcolumntype{?}{!{\vrule width 1pt}}
    \begin{tabular}{c?c|c?c|c?c|c?c|c?c|c?c|c?c|c?c|c?c|c?}
         Room & \multicolumn{2}{c?}{Balder (pretraining)} & \multicolumn{2}{c?}{Freja} & \multicolumn{2}{c?}{Munin} \\
         $\Tilde{M}$ & \multicolumn{2}{c?}{100} & $20$ & $33$ & $20$ & $33$\\
        \hline
        \multicolumn{7}{c?}{$\NMSE$ $[\si{\decibel}]$}\\
        \hline
        DP  & 
          \multicolumn{2}{c?}{-33.00}& 
        \textbf{-3.16}&  \textbf{-5.96}& 
        \textbf{-3.83} & \textbf{-7.2}\\
        FT  & 
         \multicolumn{2}{c?}{-}& 
         -2.72& -5.57 & 
         -3.82& -6.3\\
        LoRA  & 
         \multicolumn{2}{c?}{-} &
         -2.35&  -5.27& 
         -3.08& \textit{-6.9} \\
        \hline
    \end{tabular}}}
    \resizebox{0.33\textwidth}{0.05\textwidth}{%
    \centering{
    \newcolumntype{?}{!{\vrule width 1pt}}
    \begin{tabular}{c?c|c?c|c?c|c?c|c?c|c?c|c?c|c?c|c?c|c?}
         Room & \multicolumn{2}{c?}{Balder} & \multicolumn{2}{c?}{Munin (pretraining)} & \multicolumn{2}{c?}{Freja} \\
         $\Tilde{M}$ &  $20$ & $33$ & \multicolumn{2}{c?}{100} & $20$ & $33$ \\
        \hline
        \multicolumn{7}{c?}{$\NMSE$ $[\si{\decibel}]$}\\
        \hline
        DP  & 
          \textbf{-2.54} & \textbf{-5.80}& 
        \multicolumn{2}{c?}{-33.31}& 
        -3.09&  -5.82\\
        FT  & 
         -2.35& -5.78& 
         \multicolumn{2}{c?}{-} & 
         \textbf{-3.14}& \textbf{-6.42}\\
        LoRA  & 
         \textit{-2.38}& -4.55&
         \multicolumn{2}{c?}{-} & 
         -2.16&  -4.61\\
        \hline
    \end{tabular}}}
    \resizebox{0.33\textwidth}{0.05\textwidth}{%
    \centering{
    \newcolumntype{?}{!{\vrule width 1pt}}
    \begin{tabular}{c?c|c?c|c?c|c?c|c?c|c?c|c?c|c?c|c?c|c?}
         Room & \multicolumn{2}{c?}{Balder} & \multicolumn{2}{c?}{Munin} & \multicolumn{2}{c?}{Freja (pretraining)}\\
         $\Tilde{M}$ &  $20$ & $33$ & $20$ & $33$ & \multicolumn{2}{c?}{100}\\
        \hline
        \multicolumn{7}{c?}{$\NMSE$ $[\si{\decibel}]$}\\
        \hline
        DP  & 
        -2.55&  \textbf{-5.90}& 
        -3.90 & \textbf{-7.59}& 
          \multicolumn{2}{c?}{-33.22}\\
        FT  & 
         \textbf{-2.88}& -5.79 &
         \textbf{-4.46}& -7.53& 
         \multicolumn{2}{c?}{-}\\
        LoRA  & 
         -2.53&  -5.01& 
         -3.44& -6.32 &
          \multicolumn{2}{c?}{-}\\
        \hline
    \end{tabular}}}
    \vspace{1mm}
    \caption{$\NMSE$ of the considered adaptation techniques and different sampling conditions. Best values are in bold, while in italic when LoRA outperforms FT.}\vspace{-1em}
\end{table*}
In order to analyze the adaptation performance for RIR reconstruction, we adopt a MultiResUNet-based DP architecture as proposed in \cite{pezzoli2022deep}. 
MultiResUNet~\cite{ibtehaz2020multiresunet} is a CNN that extends the traditional U-Net by combining multi-resolution convolutional blocks and residual paths to extend the receptive field without increasing the overall number of parameters.
The architecture comprises 3 downsampling layers and upsampling layers with 128 filters each.
The network outputs an estimate of the RIRs $\hat{\mathbf{H}}$, while the input is a fixed random tensor $\mathbf{z} \in \mathbb{R}^{N\times M\times 128}$ of zero-mean white noise with variance $0.1$, which is sampled once and kept constant throughout training, following the DP paradigm. 

The model is implemented in PyTorch, and trained using AdamW optimizer \cite{loshchilov2017decoupled} with $0.05$ learning rate over 500 iterations.
The DP network contains approximately 2.431.876 parameters when trained from randomly initialized parameters or fine-tuned (FT), while LoRA-based fine-tuning involves significantly fewer trainable parameters depending on the selected rank $r$ in \eqref{eq:lora} with $\alpha=2r$.

The RIRs are sampled at $\SI{8}{\kilo\hertz}$ in real environments from \cite{pezzoli2021ray} and \cite{zea2019compressed}, while the inter-sensor distance within the array is $\SI{3}{\centi\meter}$. 

In order to assess the reconstruction performance we evaluate the Normalized Mean Squared Error ($\operatorname{NMSE}$) defined as \cite{zea2019compressed}
\begin{equation}\label{eq:nmse}
    \NMSE\left(\hat{\mathbf{h}}, \mathbf{h}\right) = 10\log_{10} \frac{1}{M} \sum_{m=1}^{M} \frac{\lVert \hat{\mathbf{h}}_m - \mathbf{h}_m \rVert^2}{\lVert \mathbf{h}_m \rVert^2},
\end{equation}
where $\hat{\mathbf{h}}_m\in\mathbb{R}^{N\times 1}$ and $\mathbf{h}_m\in\mathbb{R}^{N\times 1}$ are the $m$th estimate and reference RIR, respectively. 

\subsection{Scenario: Single Room}
As a first analysis, we considered a scenario where the DP is trained, using all the available $M$ channels, for a given source position $\bm{r}_1'$ (see Fig.~\ref{fig:setup_office}), and later it is adapted for a different source location $\bm{r}_2'$ within the same environment. 
During adaptation, only a randomly selected subset of microphones ($\Tilde{M}\leq M$) is employed. 
The measurements from \cite{pezzoli2021ray} consists of a ULA with $M=32$ sensors in a room with $\mathrm{T_{60}}\approx\SI{0.4}{\second}$.
In Fig.~\ref{fig:rank}, we inspect the impact of rank $r$ in \eqref{eq:lora} on the adaptation performance when all microphones are available $\Tilde{M}=32$. 
We vary $r=(1,2,4,16,32,64)$ determining a number of trainable adapter parameters \eqref{eq:lora} correspondingly to the $(0.2, 4, 7, 14, 30, 60, 122)\si{\percent}$ of the original trainable parameters in the DP. 
From Fig.~\ref{fig:rank}, we can note that the best performance is achieved for $r={16}$, while using an adapter of size around $\SI{30}{\percent}$ with respect to the overall model. 
Similar accuracy can be obtained with $r=32$. 
Interestingly, an over-parametrization ($r=64$) does not improve the performance, although it is expected to be beneficial with a larger iteration span.
In contrast, rank-1 adaptation is showing to be more effective when a very limited number of iterations (less than 100) is considered, suggesting that it would be preferable for very short adaptations.

In order to assess the impact of the available data during adaptation, we evaluate the reconstruction while varying the number of microphones $\Tilde{M}$. 
In Fig.~\ref{fig:mic}, the $\operatorname{NMSE}$ of DP trained from scratch, FT and LoRA ($r=16$) are reported as a function of the available channels. 
Starting from a pre-trained DP the reconstruction improves for all the considered cases with respect to the training of a randomly initialized network in particular for $\Tilde{M}=4$.
Excluding the case where $\Tilde{M}=32$, the performance of LoRA is close to the FT with a maximum difference of $\SI{1.6}{\decibel}$ for $\Tilde{M}=32$ although LoRA reduces by $\SI{70}{\percent}$ the number of trained parameters.
\subsection{Scenario: Multiple Rooms}
In this second scenario, we considered to pre-train DP in a given environment and adapt the model in a different room with fewer microphones $\Tilde{M}=(20, 33)$ selected as in \cite{zea2019compressed}. 
We adopted the dataset from \cite{zea2019compressed} comprising a ULA with $M=100$ channels and three rooms named Balder, Munin and Freja for which the estimated $\mathrm{T}_{30}$ are $\SI{0.32}{\second}$, $\SI{0.46}{\second}$ and $\SI{0.63}{\second}$, respectively. 
Similarly to the previous analysis, the rank of LoRA is set to $r=16$, which it has been experimentally found to be effective also in this scenario. 
In Table\,t1, the $\operatorname{NMSE}$ obtained by the different adaptation strategies is reported for every room. 
We can observe that training the DP without pretraining is able to obtain a limited improvement with respect to both FT and LoRA, in particular, when DP is pretrained over Balder room. 
This might be caused by the fact that Balder presents a lower $\mathrm{T}_{30}$ with respect to the other rooms and  both Munin and Freja share a frontal source, while Balder presents a different direction of arrival.
These aspects might make the adaptation more challenging with respect to a training from scratch. 
Nonetheless, the differences in $\operatorname{NMSE}$ between LoRA and the best results are limited in all the cases below $\SI{-1.81}{\decibel}$.
As one might expect adapting a network to a different environment with respect to the source position is more challenging due to the possible relevant differences in the acoustic characteristics of the rooms.
\section{Conclusions}\label{sec:conclusion}
In this paper we analyzed the performance of model adaptation in the context of SFR. 
In particular, we introduced the use of a LoRA for adapting a DP-based method for RIR reconstruction, and we investigated the performance in two scenarios. 
In the first case, the sound source is changed within the same environment, while in the second scenario we adapt a pre-trained DP to a different environment. 
In general, although LoRA is not able to obtain higher reconstruction accuracy with respect to FT of a previously trained DP, it achieves in-line performance with a reduction by $\approx\SI{70}{\percent}$ on the number of trainable parameters with respect to the original network. 
These results are observed in both the analyzed scenarios.
This first work suggests that adaptation strategies can be effective for RIR reconstruction and future work should explore more in detail such possibilities especially for large and supervised models.




\clearpage
\bibliographystyle{IEEEtran}
\bibliography{refs25}







\end{document}